# Exciton dynamics and exciton-phonon coupling in bulk and thin flakes of layered van der Waals antiferromagnet Ni$_2$P$_2$S$_6$

Nasaru Khan[1,*], Yuliia Shemerliuk[2], Sebastian Selter[2], Bernd Büchner[2,3], Saicharan Aswartham[2] and Pradeep Kumar[1,†]

[1] *School of Physical Sciences, Indian Institute of Technology Mandi, Mandi-175005, India*
[2] *Leibniz-Institute for Solid-state and Materials Research, IFW-Dresden, 01069 Dresden, Germany*
[3] *Institute of Solid State and Materials Physics and Würzburg-Dresden Cluster of Excellence ct.qmat, Technische Universität Dresden, 01062 Dresden, Germany*

**Abstract**

The Zhang-Rice (ZR) singlet is an intriguing quantum state offering potential to realize a spin-orbit-entangled bosonic quasiparticle, which gives rise to the Zhang-Rice exciton. Its formation is attributed to the correlation between a localized *d*-orbital of a transition metal and the *p*-orbitals of the neighbouring ligands. The layered two-dimensional (2D) antiferromagnetic Ni$_2$P$_2$S$_6$ system provide an excellent platform to probe the ZR exciton dynamics along with the role of exciton-phonon coupling. Here, we present a comprehensive study of ZR exciton and coupling with the phonons in bulk and few-layered single crystals of Ni$_2$P$_2$S$_6$ using temperature, polarization and power-dependent photoluminescence (PL) spectroscopy. At cryogenic temperatures, the PL spectra reveal distinct phonon sidebands spaced by an energy difference of ~ 117 cm$^{-1}$, indicative of exciton-phonon hybridization. Polarization-resolved measurements demonstrate a strong optical anisotropy, with a linear polarization degree of ~ 40% at 4 K. Excitation power variation highlights linear scaling of PL intensity in the low-power regime, followed by spectral deformation at higher powers attributed to the phonon-assisted recombination and exciton saturation effects. ZR exciton and phonon side bands survival temperature decreases with decreasing flake thickness suggesting their tunability. The emergence and suppression of phonon sidebands with temperature and flake thickness emphasize dimensional sensitivity and coherence limits of excitonic states. Our findings position Ni$_2$P$_2$S$_6$ as a promising candidate for tunable and anisotropic optoelectronic applications, while offering insight into quasiparticle interactions in 2D magnetic systems.

*nasarukhan736@gmail.com
†pkumar@iitmandi.ac.in

## 1. Introduction

Two-dimensional (2D) van der Waals (vdW) magnetic materials have garnered significant attention for their novel correlated phases which helps to shed light on quantum behaviours as well as device applications [1-5]. Among magnetic systems, antiferromagnets offer distinct advantages such as enhanced robustness against external perturbations, and the ability to host ultrafast spin dynamics in the terahertz regime [6,7-9]. An intricate interplay among different degrees of freedom such as spin, charge, lattice, and orbital drives the emergence of novel correlated phases in these 2D magnetic systems. Transition metal phosphorus trisulfide (TM$_2$P$_2$S$_6$) compounds represent a unique class of layered materials where both crystal and magnetic structures exhibit 2D characteristics and interlayer coupling is governed by vdW interactions.

A particularly fascinating member of this family, nickel phosphorus trisulfide (Ni$_2$P$_2$S$_6$), has recently attracted growing interest due to its rich physical phenomena, including phonon–magnon interactions, strong electron correlations, and spin-mediated excitonic effects [10-15]. Understanding the relationship between excitonic states and magnetic ordering as well as the role of phonons in these magnetically ordered materials remains a central challenge in magneto-optics, spintronics, and quantum magnetism. The van der Waals antiferromagnetic (AFM) system Ni$_2$P$_2$S$_6$ is a charge-transfer insulator with an estimated band gap of ~ 1.8 eV. Ni$^{2+}$ ions are arranged in a honeycomb lattice, each with a 3d$^8$ electronic configuration and spin (S = 1). Below the Néel temperature ($T_N$ ~ 153) K) [16-19], Ni$_2$P$_2$S$_6$ exhibits zigzag AFM ordering: magnetic moments align ferromagnetically along the crystalline *a*-axis and antiferromagnetically between neighbouring zigzag chains within the *ab*-plane. In the hexagonal network, each Ni ion is octahedrally coordinated by six S atoms with trigonal symmetry, while two P atoms are covalently bonded to six S atoms above and below the plane, forming the anionic complex complex $(P_2S_6)^{-4}$. Spins on the Ni ions are oriented either parallel or antiparallelly along the *a*-axis in the *ab*-plane, accompanied by a slight out-of-plane canting of ~ 7° [17]. The strongly correlated nature of Ni *3d* electrons plays a central role in shaping spin-related properties, including *d-d* transitions that govern the magnetic behavior. Notably, electronic Raman scattering arising from *d* orbital transitions in $Ni^{2+}$ ions is intimately tied to the spin configuration [20]. Furthermore, light-induced magnetic anisotropy has been experimentally observed by optically exciting d electrons into higher-energy orbital states of $Ni^{2+}$ [21]. Another promising development in this area is the spin-orbital coupled Zhang-Rice (ZR) state, which underscores the deep correlation between magnetism and electronic/optical degrees of freedom in these 2D magnetic systems [12, 22].

Excitonic emission in $Ni_2P_2S_6$ exhibits strong coupling with phonons, giving rise to exciton-phonon bound sidebands [14]. The microscopic origin of the exciton emission in these 2D magnetic systems, in particular $Ni_2P_2S_6$, remains an open question, with competing interpretations that include, Hund-type excitons, spin-flip-induced *d-d* transitions, spin-orbital entangled ZR excitons etc [12, 23-27]. Among these proposed mechanisms, the ZR exciton model stands out as particularly compelling, arising from electronic transitions between ZR triplet and singlet states [12, 28-29]. ZR exciton involves the complex interplay of spin, orbital, and charge degrees of freedom in strongly correlated materials. Light-matter interaction studies have shown great efficacy in exploring the fundamental properties of 2D magnetic materials, with photoluminescence (PL) spectroscopy emerging as a powerful technique for identifying characteristic optical signatures. However, PL activity is relatively weak among 2D systems in general, limiting the experimental access to their complex photophysical behaviour-particularly the hybridization of quasiparticles. TM$_2$P$_2$S$_6$ compounds stand out in this regard, as they exhibit pronounced excitonic effects driven by reduced dielectric screening in low dimensions. PL spectroscopy thus serves as a key probe in these materials, enabling insights into exciton dynamics, recombination pathways, and exciton-phonon or exciton-photon coupling mechanisms.

In the current study, we report a detailed investigation of the PL properties of Ni$_2$P$_2$S$_6$, utilizing both unpolarized and angle-resolved PL measurements. These approaches allow us to analyse comprehensively the optical behaviour, with focusing on exciton dynamics and the exciton-phonon interaction. Additionally, we explore the variation of PL response as a function of excitation laser power, offering insights into nonlinear optical phenomena and shedding light on the underlying carrier recombination processes. By probing the exciton-phonon coupling in single crystals for both bulk and thin flake geometries, this work underscores the optoelectronic potential of Ni$_2$P$_2$S$_6$ within the broader landscape of layered 2D magnetic materials.

**2. Experimental details**

Few-layer Ni$_2$P$_2$S$_6$ samples were prepared via mechanical exfoliation from bulk single crystals and transferred onto SiO$_2$/Si substrates. Flake thicknesses were characterized using Atomic Force Microscopy (AFM). Temperature-dependent PL measurements were performed on both bulk and exfoliated samples using a Horiba LabRam HR micro-Raman spectrometer integrated with a closed-cycle helium cryostat. A 633 nm (1.96 eV) laser was used as the excitation source, and a low incident power (~1 mW) was used to minimise the sample heating.

**3. Results and discussion**

**3.1. Photoluminescence of Ni$_2$P$_2$S$_6$**

Figure 1 presents the temperature-dependent PL spectra of bulk and 10 nm-thick flakes of Ni$_2$P$_2$S$_6$. We observed nine distinct peaks labelled as P1-P6, E$_A$, S$_A$, and S$_B$. Peaks P1-P6 appear prominently only at low temperatures and are attributed to the phonon sideband modes, while $S_A$ manifests as a shoulder mode. These modes remain well-resolved up to ~ 60-100 K, depending on the thickness. Among them, the $E_A$ peak stands out as a particularly sharp and intense emission centred at ~ 1.476 eV with a narrow linewidth of ~ 2 meV at 4 K. Notably, $E_A$ persists even under sub-bandgap excitation (e.g., 785 nm or 1.58 eV, data not shown), which precludes its origin from conventional inter-band transitions, considering the optical bandgap of bulk $Ni_2P_2S_6$ is ~ 1.8 eV [11]. Among the proposed mechanisms for excitonic emission in $Ni_2P_2S_6$, the ZR exciton scenario offers a compelling explanation. The ZR exciton arises from transitions between ZR singlet and triplet states, and is highly sensitive to charge redistribution and spin alignment between Ni $d$-orbitals and neighbouring S $p$-orbitals. The excitonic state emerges within a periodic array of self-doped $NiS_6$ clusters, lending support to the ZR exciton interpretation of the $E_A$ peak. Additionally, the weak shoulder mode is tentatively assigned to local inhomogeneities in the $NiS_6$ coordination environment.

To quantitatively analyse the PL spectra, we applied Lorentzian line-shape fitting to extract self-energy parameters, including peak energy, linewidth (FWHM - full width at half maximum), and the intensity. Figure 2 depicts the temperature-dependent evolution of these parameters for the $E_A$ exciton. In bulk and 20 nm-thick $Ni_2P_2S_6$, the FWHM increases monotonically with temperature. In contrast, 10 nm and 15 nm flakes exhibit a non-monotonic trend i.e. FWHM initially decreases slightly before rising with further increasing temperature. As temperature increases, the $E_A$ peak undergoes a redshift accompanied by a pronounced decrease in the intensity. In bulk sample, $E_A$ peak vanishes at ~ 120 K well below the AFM ordering temperature signalling a faster loss of excitonic coherence compared to magnetic order. This also suggests that underlying long range magnetic ordering may also be responsible for the origin of ZR exciton. Interestingly, this disappearance temperature decreases with decreasing the flake thickness; in 10 nm samples, the $E_A$ exciton persists only up to ~ 60 K.

Since AFM ordering is known to weaken progressively with reduced thickness, vanishing completely in monolayer $Ni_2P_2S_6$, the early extinction of excitonic emission in thinner flakes may be associated with the suppressed magnetic order. However, the magnitude of this temperature reduction is unexpectedly large, suggesting that additional factors beyond mere dimensionality may be contributing to exciton instability.

Figure S2 displays PL spectra in the energy range of 1.47 to 1.51 eV at 4 K, with insets showing data at 50 K, for samples ranging from bulk to 10 nm-thick $Ni_2P_2S_6$. As the thickness decreases, the $S_A$ peak becomes progressively weaker. A similar weakening of the $S_A$ peak is also observed with increasing temperature (see Fig. 1 and Supplementary Fig. S2). For 15 nm and 10 nm flakes, it vanishes at ∼ 50-60 K. The $S_B$ peak likewise diminishes with both reduced sample thickness and elevated temperature, and is absent in the 10 nm and 15 nm samples at 50 K (see Figures 1 and S2). Interestingly, in bulk samples, $S_B$ is faintly present at 4 K, strengthens up to 50 K, and subsequently fades between 60 K and 120 K. Both $S_A$ and $S_B$ modes disappear near the temperature at which the $E_A$ exciton vanishes, suggesting that these modes likely stabilize through a shared mechanism with the ZR exciton and destabilize more rapidly than the AFM order.

Figure 3 shows the PL spectra of bulk $Ni_2P_2S_6$ at 4 K, spanning the energy range of 1.34-1.50 eV. The red solid line delineates the boundary of the underlying continuum, and the inset illustrates its extraction via baseline construction from the raw data. This continuum is well described by two Gaussian components ($P_A$ and $P_B$) up to ∼ 100 K, beyond which the two features merge. The continuum resides below the phonon sidebands, with its high-energy tail overlapping with the ultra-narrow exciton peak ($E_A$), but without notably affecting the exciton's spectral shape or intensity. The continuum persists up to ∼ 270 K for bulk sample, well above the AFM transition temperature ($T_N$ ∼ 150 K), and aligns with the emergence of a broad feature previously attributed to two-magnon scattering [18, 30-31]. Notably, in 2D magnetic systems, two-magnon excitations may survive well above $T_N$ due to short-range spin correlations and pronounced quantum fluctuations [31]. As such, the continuum reflects a magnetic-order-compatible behaviour, contrasting with the sharp excitonic emission, which fades significantly even below $T_N$.

Figure 4 summarizes the temperature evolution of the $P_B$ component's peak energy, FWHM, and integrated intensity up to ~100 K. With increasing temperature, the $P_B$ peak redshifts, broadens, and weakens a trend consistent with the temperature dependence of the two-magnon mode observed near 530 cm$^{-1}$ in the Raman spectra of $Ni_2P_2S_6$ [18, 30-31]. These correlations suggest the attribution of $P_B$ to a two-magnon-related optical signature; however further experimental and theoretical studies are required to decipher their origin.

### 3.2. Temperature dependent bandgap and thermal broadening

The semiconductor bandgap exhibits temperature dependence, typically decreasing as temperature rises due to lattice thermal expansion and enhanced exciton-electron-phonon interactions. These effects collectively lead to a downward shift in band energy with increasing temperature. This temperature-driven bandgap variation is commonly described by empirical models, the most widely used being the Varshni relation [32], expressed as: $E(T) = E_0 - \dfrac{\sigma T^2}{T + \beta}$, where, $E_0$ denotes the band energy at 0 K, $\sigma$ is the temperature coefficient reflecting exciton-phonon coupling strength, and $\beta$ relates to the material's Debye temperature. Figure 2 illustrates the temperature evolution of exciton peak frequency and linewidth for both bulk and thin-flake of $Ni_2P_2S_6$. A good agreement is observed between the experimental data and the Varshni fit, represented by the red solid line. The fitting parameters extracted are summarized in Table S1.

An alternative empirical model used to describe the temperature dependence of the bandgap is provided by the O'Donnell-Chen relation [33], expressed as:

$$E(T) = E_0 - S \times E_P [\coth(\frac{E_P}{2k_B T}) - 1]$$

where, $k_B$ denotes the Boltzmann constant, $E_0$ is the excitonic band energy at 0 K, and $E_P$ represents the average phonon energy contributing to the temperature-induced shift in band energy. The dimensionless factor S, known as the Huang-Rhys parameter, quantifies the strength of exciton-electron-phonon coupling, with larger values indicating stronger coupling. The solid green line in Figure 2(a) shows the fit of the experimental data using the O'Donnell-Chen model, which yielded good agreement. The corresponding fitting parameters are summarized in Table S2.

We now analyse the temperature-dependent evolution of the exciton linewidth. The linewidth is significantly influenced by interactions between excitons and longitudinal optical (LO) phonons, acoustic phonons, as well as free carriers. Additionally, contributions from intrinsic defects and extrinsic impurities further broaden the excitonic features. Figure 2(b) illustrates the temperature-driven changes in the linewidth for both bulk and few-layer $Ni_2P_2S_6$. Across all systems, the excitonic band exhibits homogeneous broadening with increasing temperature, underscoring the role of exciton-phonon coupling. At finite temperatures, the linewidth of an excitonic transition in semiconductors can be described by the following expression [34]:

$$\Gamma(T) = \Gamma_0 + \Gamma_{AC}(T) + \Gamma_{LO}(T) = \Gamma_0 + \lambda_{AC} T + \lambda_{LO} N_{LO}(T)$$

Here, the first term $\Gamma_0$ represents the linewidth at 0 K, primarily arising from exciton scattering due to impurities and lattice imperfections. The second and third terms correspond to the linewidth contributions induced by exciton-acoustic phonon and exciton-LO phonon interactions, respectively. The coefficients $\lambda_{AC}$ and $\lambda_{LO}$ denote the coupling strengths associated with exciton-acoustic phonon and exciton-LO phonon scattering mechanisms. The interaction between excitons and LO phonons is governed by the Fröhlich mechanism [35], whereas exciton coupling with acoustic phonons can be understood via the deformation potential model [36]. The phonon population at finite temperature is described by the Bose-Einstein distribution function, given by $N_{LO}(T) = 1/[e^{E_{LO}/k_B T} - 1]$, where $E_{LO}$ is the energy of the interacting LO phonons.

The solid blue line in Figure 2(b) shows a fit to the experimental data using the above expression, which yields a very good agreement across the temperature range. The extracted fitting parameters are summarized in Table S3. Figures 5(a)-(c) further isolate the individual contributions of acoustic and LO phonons to the excitonic linewidth for bulk, 20 nm, and 15 nm $Ni_2P_2S_6$ samples, respectively. At low temperatures (below ~70 K), both acoustic and LO phonons contribute to linewidth broadening. However, acoustic phonons dominate due to their lower energy compared to thermal energy, resulting in a higher population and stronger exciton-acoustic phonon scattering. In contrast, LO phonons have relatively higher energies and remain sparsely populated in this regime, contributing less to the broadening. Above ~ 70 K, the thermal energy surpasses the LO phonon energy, leading to an increased LO phonon population and a marked rise in linewidth broadening via exciton-LO phonon interactions. This transition marks the regime where LO phonon scattering becomes the dominant mechanism. Notably, the contribution from LO phonons remains nearly constant up to ~ 40 K and exhibits a linear temperature dependence beyond this point in all systems. Furthermore, the disparity between acoustic and LO phonon-induced broadening is less pronounced in bulk $Ni_2P_2S_6$ compared to few-layer samples, as evident in Figure 5.

## 3.3. Polarization dependence of PL

Polarization-dependent PL measurements can be performed using several configurations, involving the variation of polarization of either the incident or scattered light, or through sample rotation. In the present study, we did polarization-dependent measurements using two distinct configurations. In the first configuration, the polarization of the incident laser was rotated while maintaining a fixed orientation for the scattered light. The resulting PL spectra, shown in Figure 6(a, c), revealed no significant polarization dependence in PL intensity. In the second configuration, the scattered light's polarization was varied while the incident light direction was fixed. This setup revealed a pronounced polarization dependence: the PL intensity exhibited maxima when the polarization direction was aligned parallel to the crystallographic *a*-axis, and minima when perpendicular (see Fig. 6 (b) and (d)). The lack of polarization dependence when rotating the incident light, contrasted with the dependence when rotating scattered light, suggesting a directional preference in emission mechanisms rather than absorption. This may be due to anisotropic dipole orientation in the excitonic transitions, likely due to crystal field effects or orbital symmetry along the crystallographic axes.

The angular dependence of PL intensity was modelled using a sinusoidal function $I(\theta) = I_0 + I_A \cos^2 \theta$, where $\theta$ is the angle between the PL polarization direction and the *a*-axis of $Ni_2P_2S_6$, and $I_0$ and $I_A$ are the fitting parameters. To further quantify the polarization behaviour, we evaluated the linear degree of polarization of the excitonic peak. It is defined as $\delta = (I_b - I_a)/(I_b + I_a)$ [13], where $I_b$ and $I_a$ are PL intensities measured with scattered light polarized along the *b*-axis and *a*-axis, respectively. At 4 K, the linear polarization degree for $Ni_2P_2S_6$ was found to be ~ 40%. A linear polarization degree of ~ 40% is substantial suggesting strong polarized excitonic emission, which may be valuable for polarization-sensitive optoelectronic applications, such as quantum emitters or anisotropic light modulators in layered systems.

### 3.4. Excitation power dependence of PL

PL spectra of bulk Ni₂P₂S₆ were also recorded by systematically varying the excitation laser power at different temperatures. Figure 7(a) presents the PL spectra at 4 K for a range of excitation powers. As the power increases from ~ 86 µW to ~ 0.8 mW, the intensity of the excitonic peak exhibits an approximately linear enhancement, consistent with the expected behaviour under low excitation conditions. However, upon further increasing the power to ~ 4.25 mW, a marked change in the excitonic feature is observed. At this high power, the excitonic peak evolves into a shoulder-like feature, while phonon sidebands become significantly more pronounced. The near-linear increase in excitonic PL intensity up to ~ 0.8 mW reflects a regime dominated by radiative recombination of unperturbed excitons, with minimal phonon scattering or non-radiative losses. At higher power in the non-linear regime, emergence of shoulder-like features and stronger phonon sidebands may be due to (i) Exciton saturation where the recombination rate plateaus due to limited available states. (ii) Enhanced phonon-assisted processes as stronger laser excitation may couple excitons more efficiently to lattice vibrations; and (iii) Local heating may activate additional phonon modes or modify bandgap energies subtly.

Figure 7(b) presents the evolution of integrated PL intensity with excitation laser power at three temperatures (4K, 30K and 80K). Within the power range of ~ 0.086 - 0.85 mW, the PL intensity exhibits an approximately linear increase with incident power. To quantitatively analyse this behaviour, we fitted the integrated intensity of the excitonic peak using a generalized power-law function $I = \eta P^{\alpha}$, where I is intensity, P is the excitation power, η is a proportionality constant, and α is the power exponent. The coefficient $\eta$ encapsulates contributions from absorption, carrier capture, and recombination processes [37-38], while $\alpha$ reflects the nature of the recombination mechanism which is unity for excitonic recombination and ~ 2 for free carrier recombination. The extracted values of $\eta$ and $\alpha$ are summarized in Table S4. Temperature-dependent variations in $\alpha$ offer insight into recombination dynamics.

At 30 K and 80 K, $\alpha$ values of ~ 0.6 and ~ 0.7 suggest predominant radiative recombination. In contrast, a higher value of ~ 1.1 at 4 K points to the emergence of nonradiative channels, possibly due to increased exciton-defect or exciton-phonon scattering. Furthermore, for pure radiative processes, $\alpha$ is expected to remain below unity, since the higher excitation powers tend to induce optical dissipation through light diffusion and nonradiative carrier losses [37]. Consistent with PL quenching, the coefficient $\eta$ demonstrates a decreasing trend with rising temperature (from 4 K to 80 K), further confirming temperature-induced suppression of excitonic emission.

In addition, we extracted self-energy parameters from the PL spectra to understand the relationship between excitonic recombination dynamics in $Ni_2P_2S_6$. Figures 7(c) and 7(d) display the evolution of peak energy and FWHM for the sharp ZR exciton as a function of excitation power. Notably, both parameters remain nearly invariant with increasing power at low power and decreases/increases slightly with power above ~ 0.5 mW, indicating that the excitonic state is stable and not significantly perturbed under moderate excitation.

### 3.5. Exciton-phonon coupling

Figure 8(a) presents the fitted PL spectrum acquired at 4 K for the bulk $Ni_2P_2S_6$. Remarkably, several satellite peaks appear on the low-energy side of the primary $E_A$ peak, resembling phonon sidebands. These sidebands, which exhibit recurring phonon modes, are characteristic signatures of quasiparticle hybridization-most notably exciton-phonon or magnon-phonon interactions. To resolve the spectral features quantitatively, the PL spectrum was fitted using Lorentzian line shape functions, allowing precise determination of the energy spacing between peaks. The fitting analysis revealed that the satellite peaks are spaced by an energy difference of ~ 117 ± 8 cm$^{-1}$ (0.0146 eV), suggesting the involvement of phonons in the PL process. Comparison with the Raman spectrum of $Ni_2P_2S_6$ confirms that this shift aligns closely with the energy of the $A_g$ phonon mode at ~ 131 cm$^{-1}$, associated with the vibration of heavy Ni metal ions. Accordingly, the low-energy peaks labelled as P1-P6 are attributed to the phonon sidebands arising from exciton-phonon coupling. The peak energies of these sidebands remain largely unchanged with varying excitation power (see Figure 8(e)).

Figure 8(b) shows the excitation power dependence of PL intensity for peaks P2, P3, P4, and P6. Each follows a generalized power-law relationship $I = \eta P^\alpha$. Notably, the PL intensities exhibit near-linear scaling with excitation power. The exponent $\alpha$, which offers insight into recombination mechanisms, is found to be ~ 1.2 ± 0.1, at 4 K for all investigated peaks. This suggests the emergence of nonradiative channels even at low temperature, contributing to exciton-phonon coupling dynamics. Taken together, the consistent power-law exponents and invariant peak positions further support the assignment of P1-P6 as phonon sidebands linked to hybrid excitonic states in $Ni_2P_2S_6$. To further corroborate the excitonic origin of the observed sidebands in the PL spectrum, we investigated the temperature-dependent evolution of peak positions for P2, P3, P4 and P6, as shown in Figure 8(d). Evidently, the change in peak position as a function of temperature shows a persistent behaviour for all four peaks, providing support for the hypothesis of having the same origin of these peaks. Additionally, Figure 8(c) illustrates the peak positions measured across different flake thicknesses of $Ni_2P_2S_6$. The near-invariance of these energies with respect to sample thickness further supports the attribution of these peaks to exciton-phonon coupling processes intrinsic to the material.

To further investigate the temperature dependence of exciton-phonon coupling dynamics, we performed temperature dependent PL measurements on $Ni_2P_2S_6$. Figure 1(a) shows the evolution of the PL spectrum for bulk $Ni_2P_2S_6$ as the temperature increases from 4 K to 270 K. Within the 4-80 K range, distinct and well-resolved phonon sidebands are observed in the bulk sample, indicating robust exciton-phonon interactions. In contrast, for 15 nm (not shown here) and 10 nm thinner flakes samples these phonon sidebands are observable only up to ~ 50 K. As temperature rises further, these sidebands progressively lose their individuality and eventually merge into two broader peaks. This spectral merging may be attributed to thermal broadening of the phonon energy distribution [39], which blurs individual phonon contributions and modifies the emission landscape.

Supplementary Figure S3 highlights the exciton-phonon coupling branches in the baseline-corrected PL spectrum recorded at 4 K. To investigate the temperature-dependent dynamics of these coupling processes in $Ni_2P_2S_6$, we analysed the peak energy, FWHM, and integrated PL intensity of the exciton-phonon branch, as shown in Figure 9. The peak energy exhibits a consistent red shift as temperature increases from 4 K to 80 K, a characteristic behaviour commonly observed in semiconductors [32, 39]. This red shift arises from temperature-dependent electron-lattice interactions. Additionally, thermal expansion of the lattice contributes to the narrowing of the band gap with increasing temperature. Figure 9(c) demonstrates that the PL intensity of the exciton-phonon branch decreases with rising temperature, which again reflects expected semiconductor behaviour. In $Ni_2P_2S_6$, PL intensity is predominantly governed by exciton generation and radiative recombination, modulated by factors such as exciton binding energy, magnetic ordering, and the presence of nonradiative recombination pathways. The increase in PL intensity at lower temperatures can be attributed to suppressed thermal escape of photoexcited carriers [40]. FWHM of the PL spectrum follows typical temperature dependence, broadening with increasing temperature due to enhanced phonon scattering and reduced exciton coherence.

**Conclusion**

In this work, we have comprehensively explored exciton-phonon coupling dynamics in bulk and few-layer $Ni_2P_2S_6$ through in-depth temperature, polarization and power-dependent PL spectroscopy. The emergence of distinct phonon sidebands in low-temperature PL spectra, spaced by well-defined Raman shifts, provides compelling evidence for exciton-phonon hybridization. Power-dependent analyses further revealed linear scaling in PL intensity across low excitation powers, followed by spectral reshaping at higher powers a hallmark of phonon-assisted recombination and potential exciton saturation phenomena. We observed a sharp emission centred at ~ 1.47 eV attributed to the Zhang-Rice exciton. Our polarization dependent PL measurements reveal the significant degree of linear polarization for sharp ZR exciton. The polarization-resolved PL measurements underscore the intrinsic optical anisotropy of $Ni_2P_2S_6$, with emission strongly modulated by the orientation of scattered light, leading to a linear polarization degree of ~ 40% at 4 K. Taken together, our findings not only affirm the robust excitonic character and phonon interactions in $Ni_2P_2S_6$ but also highlight its potential for anisotropic, tunable optoelectronic applications. The persistence of phonon sidebands and excitonic coherence at cryogenic temperatures makes this material a compelling platform for studying quasiparticle dynamics in 2D magnetic systems.

**Conflicts of interest**

There are no conflicts of interest to declare.

**Acknowledgement**

NK acknowledge CSIR India for the fellowship. PK acknowledge support from IIT Mandi for the experimental facilities and ANRF (CRG/2023/002069) for the financial support. SA acknowledges Deutsche Forschungsgemeinschaft (DFG) through Grant No. AS 523/4–1 and BB through SFB 1143 (project-id 247310070), ct.qmat (EXC 2147, project-id 390858490).

**Data availability statement**

All data that supports the findings of this study are included within the article and supplementary file.

**References**

[1] B. Huang, G. Clark, E. Navarro-Moratalla, D. R. Klein, R. Cheng, K. L. Seyler, D. Zhong, E. Schmidgall, M. A. McGuire, D. H. Cobden, W. Yao, D. Xiao, P. Jarillo-Herrero, X. Xu, Nature, **546**, 270, (2017).

[2] C. Gong, L. Li, Z. Li, H. Ji, A. Stern, Y. Xia, T. Cao, W. Bao, C. Wang, Y. Wang, Z. Q. Qiu, R. J. Cava, S. G. Louie, J. Xia, X. Zhang, Nature, **546,** 265, (2017).

[3] K. S. Burch, D. Mandrus, J.-G. Nature, **563**, 47, (2018).

[4] C. Gong, X. Zhang, Science, **363**, 4450, (2019).

[5] B. Huang, M. A. McGuire, A. F. May, D. Xiao, P. Jarillo-Herrero, X. Xu, Nat. Mater., **19**, 1276, (2020).

[6] M. Mi, H. Xiao, L. Yu, Y. Zhang, Y. Wang, Q. Cao, Y. Wang, Mater. Today Nano, **24**, 100408, (2023).

[7] S. Rahman, J. F. Torres, A. R. Khan, Y. Lu, ACS Nano, **15**, 17175, (2021).

[8] V. Baltz, A. Manchon, M. Tsoi, T. Moriyama, T. Ono, Y. Tserkovnyak, Rev. Mod. Phys. , **90**, 015005, (2018).

[9] A.V. Kimel, A. Kirilyuk, A. Tsvetkov, R.V. Pisarev, T. Rasing, Nature, **429**, 850**,** (2004).

[10] N. Khan et al., 2D Mater., **11**, 035018 (2024).

[11] S. Y. Kim, T. Y. Kim, L. J. Sandilands, S. Sinn, M.-C. Lee, J. Son, S. Lee, K. Y. Choi, W. Kim, B.-G. Park, C. Jeon, H.-D. Kim, C.-H. Park, J.-G. Park, S. J. Moon, T. W. Noh, Phys. Rev. Lett., **120**, 136402, (2018).

[12] S. Kang, K. Kim, B. H. Kim, J. Kim, K. I. Sim, J. U. Lee, S. Lee, K. Park, S. Yun, T. Kim, A. Nag, A. Walters, M. Garcia-Fernandez, J. Li, L. Chapon, K. J. Zhou, Y. W. Son, J. H. Kim, H. Cheong, J. G. Park, Nature, **583**, 785, (2020).

[13] X. Wang, J. Cao, Z. Lu, A. Cohen, H. Kitadai, T. Li, Q. Tan, M. Wilson, C. H. Lui, D. Smirnov, S. Sharifzadeh, X. Ling, Nat. Mater., **20**, 964 (2021).

[14] K. Hwangbo, Q. Zhang, Q. Jiang, Y. Wang, J. Fonseca, C. Wang, G. M. Diederich, D. R. Gamelin, D. Xiao, J.-H. Chu, W. Yao, X. Xu, Nat. Nanotechnol., **16**, 655, (2021).

[15] J. H. Lee et al., Adv. Funct. Mater, **2405153**, 1, (2024).

[16] P. A. Joy, S. Vasudevan, Phys.Rev.B., **46**, 5425, (1992).

[17] A. R. Wildes, V. Simonet, E. Ressouche, G. J. Mcintyre, M. Avdeev, E. Suard, S. A. J. Kimber, D. Lançon, G. Pepe, B. Moubaraki, T. J. Hicks, Phys. Rev. B., **92**, 224408 (2015).

[18] K. Kim, S. Y. Lim, J.-U. Lee, S. Lee, T. Y. Kim, K. Park, G. S. Jeon, C.-H. Park, J.-G. Park, H. Cheong, Nat. Commun., **10**, 345 (2019).

[19] R. Basnet, A. Wegner, K. Pandey, S. Storment, J. Hu, Phys. Rev. Mater., **5**, 064413 (2021).

[20] X. Wang et al., Sci. Adv., **8**, eabl7707 (2022).

[21] D. Afanasiev, J. R. Hortensius, M. Matthiesen, S. Mañas-Valero, M. Šiškins, M. Lee, E. Lesne, H. S. J. van der Zant, P. G. Steeneken, B. A. Ivanov, E. Coronado, A. D. Caviglia, Sci. Adv. ,**7**, eabf3096, (2021).

[22] F. C. Zhang and T. M. Rice, Phys. Rev. B, **37**, 3759 (1988).

[23] C. A. Belvin et al., Nat. Commun., **12**, 4837 (2021).

[24] W. He et al., Nat. Commun., **15**, 3496 (2024).

[25] D. Jana et al., Phys. Rev. B, **108**, 115149 (2023).

[26] T. Klaproth et al., Phys. Rev. Lett., **131**, 256504 (2023).

[27] A. Shcherbakov et al., ACS Nano, **17**, 10423 (2023).

[28] F. Song et al., Nat. Commun., **15**, 7841 (2024).

[29] X. Wang et al., Nat. Commun., **15**, 8011 (2024).

[30] S. Rosenblum et al., Phys. Rev. B, **49**, 4352 (1994).

[31] S. Rosenblum and R. Merlin, Phys. Rev. B, **59**, 6317 (1999).

[32] Y. P. Varshni, Physica, **34**, 149 (1967).

[33] K. P. O'Donnell and X. Chen, App. Phys. Lett., **58**, 2924 (1991).

[34] S. Rudin et al., Phys. Rev. B: Condens. Matter Mater. Phys., **42**, 11218 (1990).

[35] R. M. Martin and T. C. Damen, Phys. Rev. Lett., **26**, 86 (1971).

[36] D. Kumar et al., J. Phys.: Condens. Matter, **32**, 415702 (2020).

[37] S. Martini, A.A. Quivy, A. Tabata, Leite R, J Appl Phys, **90**, 2280 (2001).

[38] S. Jin, Y. Zheng, A. Li, J Appl Phys, **82**, 3870, (1997).

[39] G. Antonius, S. G. Louie, Phys. Rev. B, **105**, 085111(2022).

[40] M. Kondo, N. Okada, K. Domen, K. Sugiura, C. Anayama, T. Tanahashi, J. Electron. Mater. **23**, 355 (1994).

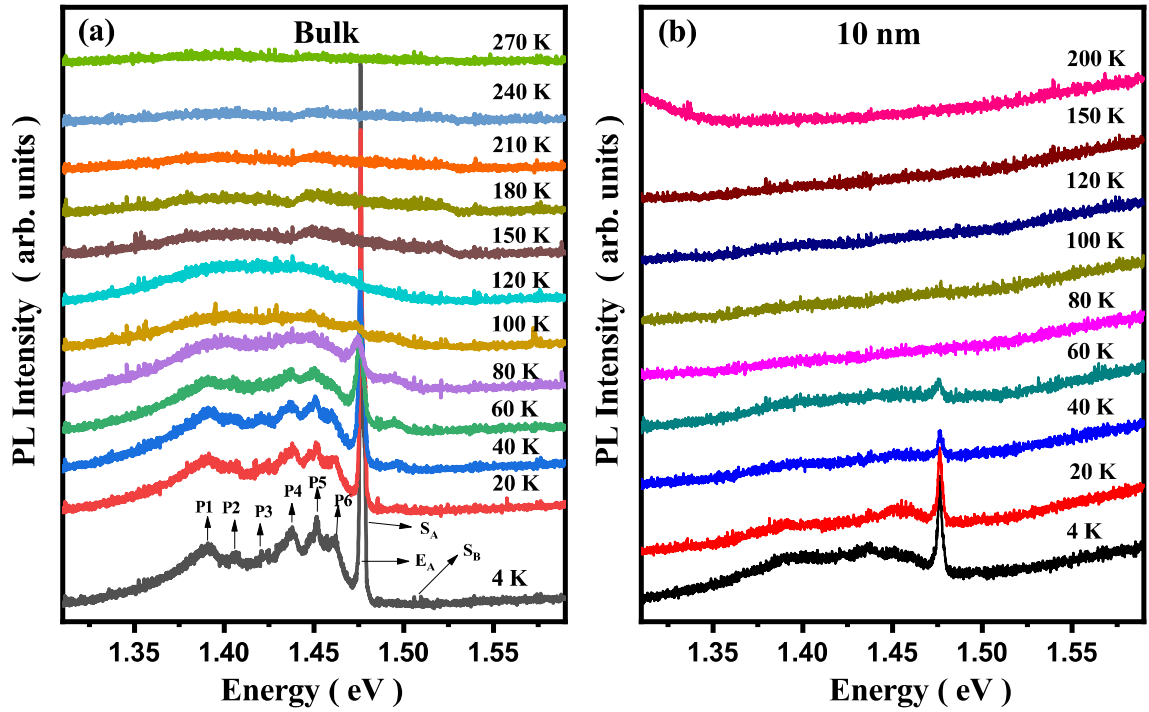

**Figure 1.** Temperature-dependent photoluminescence spectra for Ni$_2$P$_2$S$_6$. (a) Bulk sample and (b) 10 nm-thick flake, illustrating spectral evolution with increasing temperature.

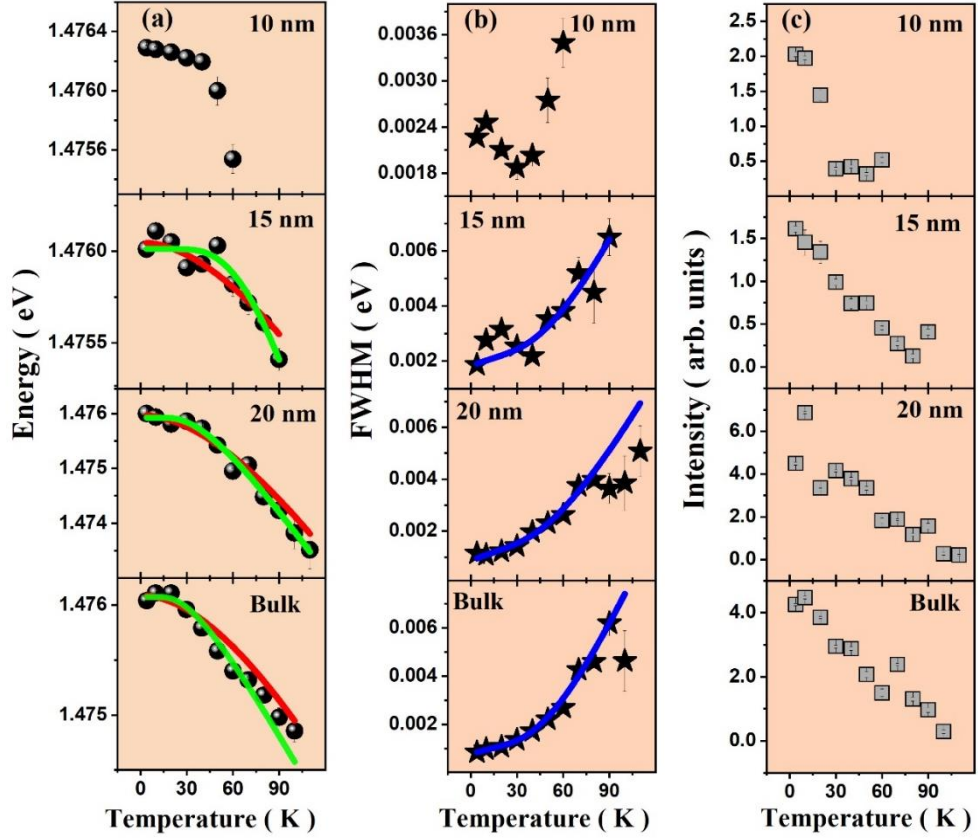

**Figure 2.** Temperature-dependent evolution of excitonic peak $E_A$ parameters in bulk and layered $Ni_2P_2S_6$ systems. Displayed plots show variations in (a) peak frequency, (b) full width at half maximum (FWHM), and (c) integrated intensity. Red and green solid lines represent fits to the peak frequency using the Varshni and O'Donnell-Chen models, respectively. The blue solid line denotes the FWHM fit based on the exciton-phonon coupling model discussed in the text.

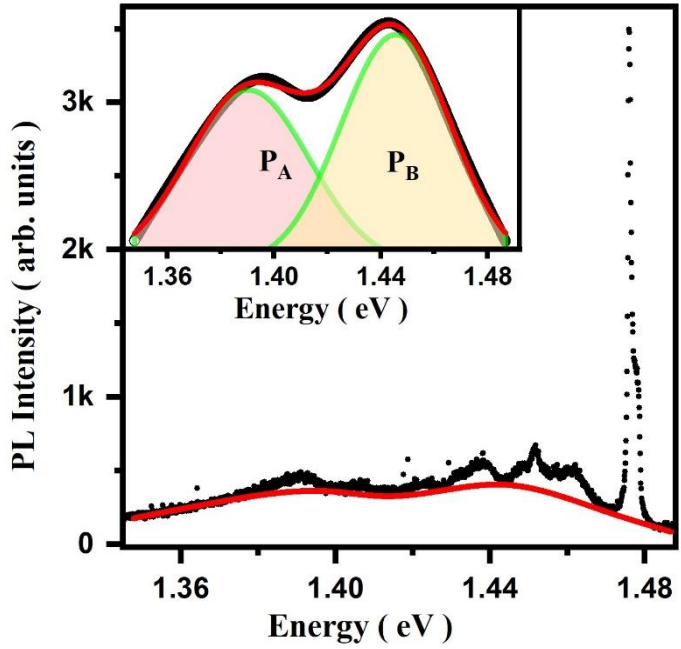

**Figure 3**. Photoluminescence spectra of Ni$_2$P$_2$S$_6$ at 4 K. The red solid line highlights the magnetic continuum contribution within the PL spectrum. The inset displays the fitted background continuum, modelled using a Gaussian line shape function.

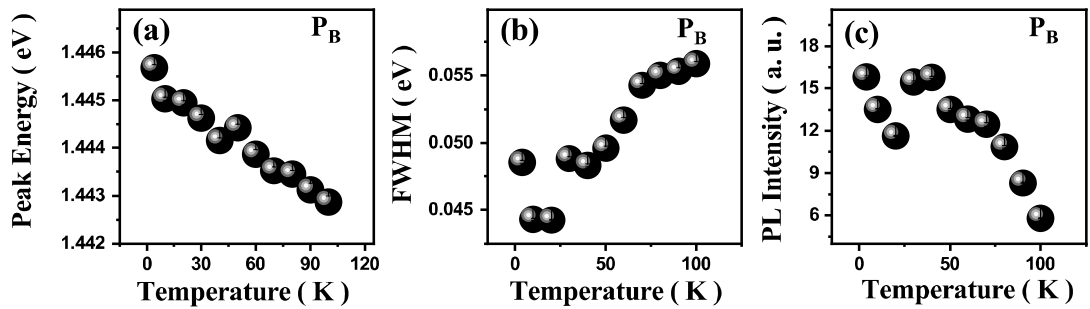

**Figure 4**. Temperature-dependent behaviour of the continuum (peak P$_B$) in Ni$_2$P$_2$S$_6$. Plots depict the evolution of (a) peak energy, (b) FWHM, and (c) integrated PL intensity of peak P$_B$ as a function of temperature.

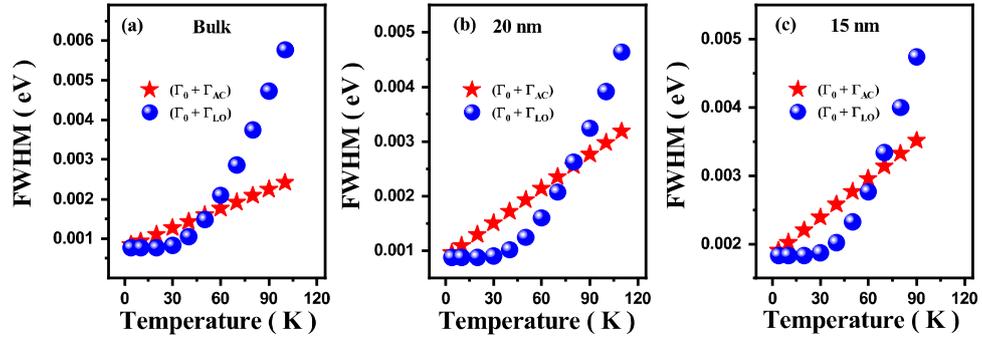

**Figure 5.** Individual phonon contributions to the linewidth of excitonic peak $E_A$ in $Ni_2P_2S_6$. (a) Bulk sample, (b) 20 nm-thick flake, and (c) 15 nm-thick flake, illustrating the respective roles of acoustic and longitudinal optical (LO) phonons in linewidth broadening.

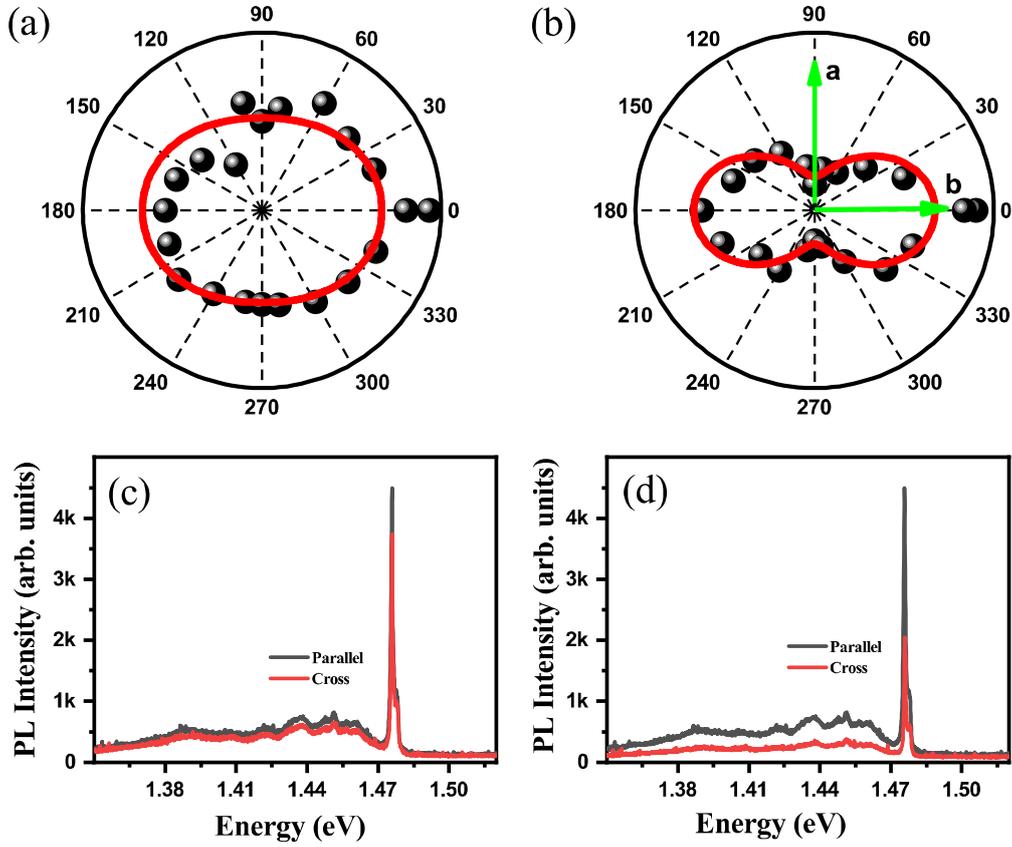

**Figure 6.** Polarization-resolved PL intensity of excitonic mode $E_A$ at 4 K. (a) PL intensity as a function of incident laser polarization; (b) PL intensity as a function of scattered light polarization. Crystal orientations (*a*- and *b*-axes) are indicated by green arrows. (c/d) PL spectra under parallel (black) and cross-polarization (red) configurations for incident/scattered laser light polarization.

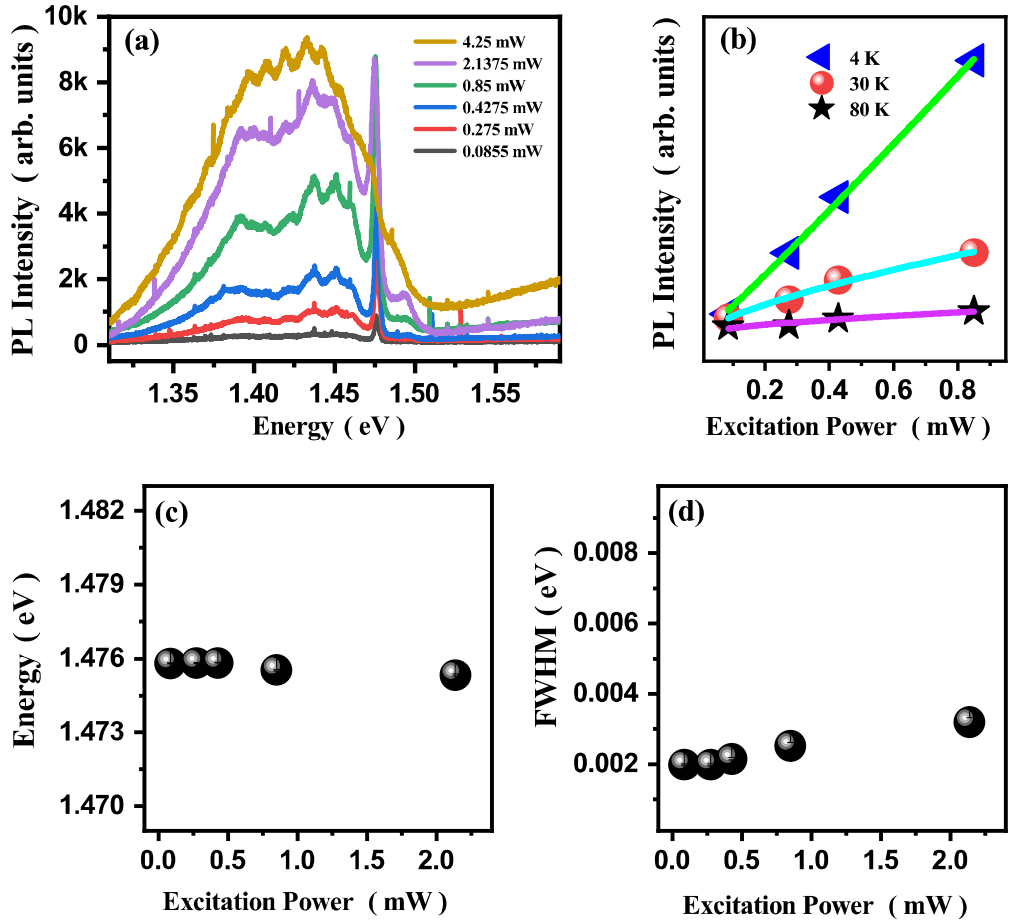

**Figure 7.** Excitation power-dependent PL spectra of bulk $Ni_2P_2S_6$. (a) PL spectra recorded at 4 K for varying excitation powers. (b) Integrated PL intensity of the Zhang-Rice excitonic peak plotted as a function of excitation power at 4 K (blue triangles), 30 K (red spheres), and 80 K (black stars). Solid lines correspond to the power-law fits as described in the text. (c) Peak energy and (d) FWHM of the ZR exciton at 4 K as a function of excitation power.

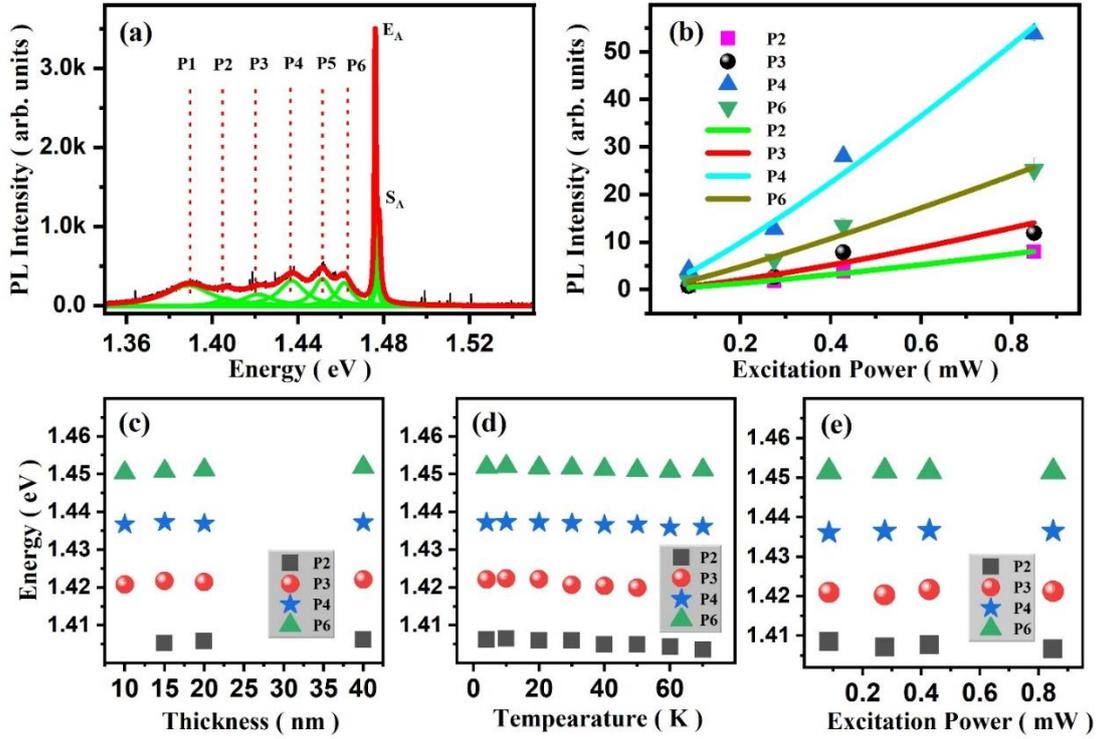

**Figure 8.** Analysis of exciton-phonon coupled modes in $Ni_2P_2S_6$. (a) Fitted PL spectrum of bulk $Ni_2P_2S_6$ at 4 K, highlighting phonon sidebands labelled as P1-P6. (b) Power-dependent PL intensity of peaks P2, P3, P4 and P6 at 4 K; solid lines represent power-law fits as described in the text. (c) Variation in peak energy of P2, P3, P4 and P6 as a function of flake thickness. (d) Temperature-dependent evolution of peak energies of P2, P3, P4 and P6 for bulk $Ni_2P_2S_6$. (e) Excitation power dependence of peak energies of P2, P3, P4 and P6 in bulk $Ni_2P_2S_6$.

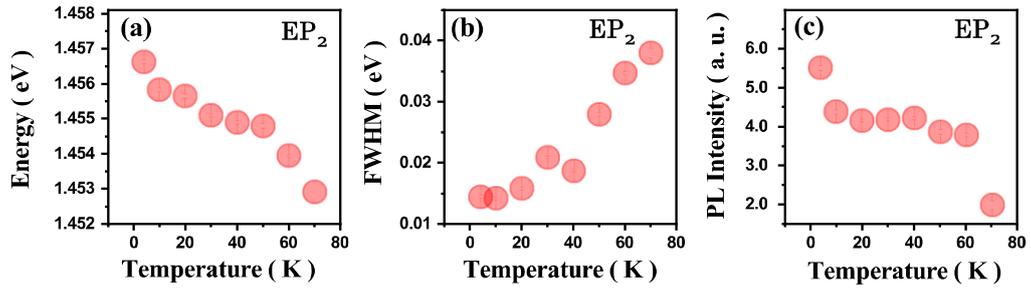

**Figure 9.** Temperature evolution of (a) peak energy (b) FWHM and (c) PL intensity for the exciton-phonon coupling branch.

# Supplementary Information

# Exciton dynamics and exciton-phonon coupling in bulk and thin flakes of layered van der Waals antiferromagnet $Ni_2P_2S_6$

Nasaru Khan[1,*], Yuliia Shemerliuk[2], Sebastian Selter[2], Bernd Büchner[2,3], Saicharan Aswartham[2] and Pradeep Kumar[1,†]

[1] *School of Physical Sciences, Indian Institute of Technology Mandi, Mandi-175005, India*
[2] *Leibniz-Institute for Solid-state and Materials Research, IFW-Dresden, 01069 Dresden, Germany*
[3] *Institute of Solid State and Materials Physics and Würzburg-Dresden Cluster of Excellence ct.qmat, Technische Universität Dresden, 01062 Dresden, Germany*

[*]nasarukhan736@gmail.com
[†]pkumar@iitmandi.ac.in

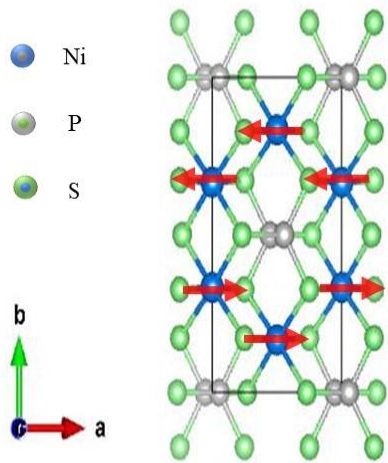

**Figure S1.** Top view of crystal structure of $Ni_2P_2S_6$. Red arrows show the zigzag antiferromagnetic spin configuration.

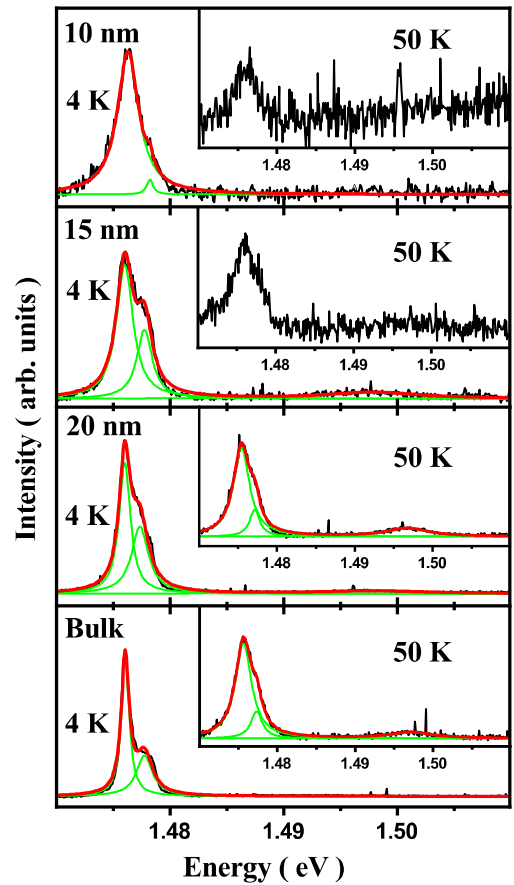

**Figure S2.** Fitted PL spectra for (a) 10 nm $Ni_2P_2S_6$ (b) 15 nm $Ni_2P_2S_6$ (c) 20 nm $Ni_2P_2S_6$ (d) Bulk $Ni_2P_2S_6$ at 4K. Insets show the fitted (for bulk and 20 nm) and raw (for 15 and 10 nm) PL spectra at 50K. The solid red lines are the total sum of Lorentzian fit to the experimental data and green lines are the individual fits of modes.

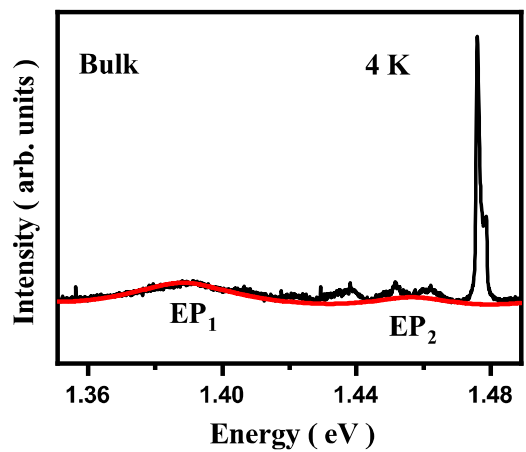

**Figure S3.** Baseline corrected PL spectra of bulk $Ni_2P_2S_6$ showing the exciton-phonon branches $EP_1$ and $EP_2$.

**Table S1.** Fitted parameters obtained using Varshni relation as described in text.

| Parameter | Bulk $Ni_2P_2S_6$ | 20 nm $Ni_2P_2S_6$ | 15 nm $Ni_2P_2S_6$ |
|---|---|---|---|
| $E_0(meV)$ | 1476.09 ± 0.02 | 1475.97 ± 0.02 | 1476.05 ± 0.03 |
| $\sigma(10^{-5}eVK^{-1})$ | 3.78 ± 0.72 | 5.99 ± 0.82 | 2.00 ± 0.74 |
| $\beta(K)$ | 230.34 ± 34.39 | 225.38 ± 32.26 | 234.86 ± 36.86 |

**Table S2.** Fitted parameters obtained using Donnell and Chain relation as described in text

| Parameter | Bulk $Ni_2P_2S_6$ | 20 nm $Ni_2P_2S_6$ | 15 nm $Ni_2P_2S_6$ |
|---|---|---|---|
| $E_0(meV)$ | 1476.07 ± 0.02 | 1475.93 ± 0.02 | 1476.01 ± 0.03 |
| $S$ | 0.15 ± 0.05 | 0.24 ± 0.06 | 0.24 ± 0.09 |
| $E_P(meV)$ | 8.51 ± 2.54 | 10.63 ± 2.52 | 22.85 ± 9.82 |

**Table S3.** Fitted parameters obtained using the expression as described in text

| Parameter | Bulk $Ni_2P_2S_6$ | 20 nm $Ni_2P_2S_6$ | 15 nm $Ni_2P_2S_6$ |
|---|---|---|---|
| $\Gamma_0(meV)$ | 0.77 ± 0.02 | 0.88 ± 0.03 | 1.83 ± 0.13 |
| $\lambda_{AC}(\mu eVK^{-1})$ | 16.40 ± 2.18 | 20.85 ± 3.14 | 18.69 ± 8.98 |
| $\lambda_{LO}(meV)$ | 25.38 ± 11.33 | 18.94 ± 4.64 | 20.52 ± 13.29 |
| $E_{LO}(K)$ | 180.37 ± 29.93 | 197.86 ± 34.99 | 187.75 ± 32.69 |

**Table S4.** Fitted parameters obtained using Power law fitting as described in text.

| | 4 K | 30 K | 80 K |
|---|---|---|---|
| $\alpha$ | 1.10 | 0.70 | 0.60 |
| $\eta$ | 30.86 | 8.76 | 2.38 |